# Design of racetrack ring resonator based dielectric laser accelerators

Huiyang Deng,[1,a,b] Jiaqi Jiang,[2,b] Yu Miao,[1] Kenneth J. Leedle,[1] Hongquan Li,[1] Olav Solgaard,[1] Robert L. Byer,[3] and James S. Harris[1,3]

[1]*Department of Electrical Engineering, Stanford, California, 94305, USA*
[2]*Department of Physics, Tsinghua University, Beijing, 100084, China*
[3]*Department of Applied Physics, Stanford, California, 94305, USA*

In this paper, we propose a novel design of dielectric laser-driven accelerator (DLA) utilizing evanescent electric field of racetrack ring resonator structures. Driven by laser light with the correctly designed optical phase window, sustained acceleration of electrons with controlled deflection is shown. Based on this design, we calculate an acceleration from 30 keV to 148.312 keV in 104.655 μm using a cascaded 11-stage racetrack ring resonators. This new idea poses a solution for on-chip integration of many DLA stages, while maintains high average accelerating gradients, providing a potential practical realization for "accelerator on a chip".

High-energy particle beams generated by high power radio frequency (RF) accelerator facilities have been foundational elements in fundamental scientific research,[1-3] biomedical applications[4,5] and coherent X-ray generation[6] for the past seven decades. Unfortunately, their widespread availability is severely limited by their size and cost, spawning a need for alternative accelerators operating in regimes outside of RF. Working at optical frequencies, dielectric laser accelerators (DLAs) – transparent laser-driven photonic structures whose near fields can synchronously accelerate charged particles – have recently demonstrated high-gradient acceleration with a variety of laser wavelengths, materials, and electron beam parameters.[7-18] The first successful DLA demonstrations were based on inverse Smith-Purcell gratings with single sided laser illumination, in which an mode that exponentially decays from the surface is excited and synchronously accelerates electrons.[7-9, 12, 13] However, this mode also exerts a phase dependent deflection force on electrons,[10, 11] requiring more studies on additional focusing and beam steering elements for sustained long-distance acceleration. In addition, most of the work done in DLA so far is still single-stage acceleration with one laser spot interacting with electrons. The only two-stage acceleration work reported so far involved challenges in aligning the lasers to the micro-size DLA structures, and is difficult to be scaled up for more stages.[15-17] To achieve a practical multi-stage "accelerator on a chip" system, it is necessary to have accelerator components and wave guiding structures integrated monolithically.

In this letter, we present a novel DLA design based on racetrack ring resonators, which enables monolithic integration and at the same time provides very good acceleration performance: high accelerating gradient (ultimately limited by laser induced material breakdown) with very high field ratio (defined in Equation (1)), controlled deflection, sustained energy gains and lower laser input demand. While racetrack ring resonators have played a significant role in photonics and various applications[19-24] but are proposed here as a concept for an electron accelerator.

A schematic of the proposed racetrack ring resonator accelerator (RRRA) structure is shown in Fig. 1(a). The waveguides and double racetrack ring resonators are Si ridges on a SiO$_2$ substrate, which could be fabricated from a silicon-on-insulator (SOI) wafer, with the structure layout symmetric to the middle channel through which the sub-relativistic electrons travel. We propose to use a 1 Hz 10 ps pulse laser at 2 μm wavelength as the power source for the laser-driven electron accelerator. We performed finite-difference-time-domain (FDTD) simulations with commercial software.[25] We find the fundamental mode of the bus waveguide, and denote the peak electric field amplitude of the mode as $E_{\text{inc}}$. This mode is then injected from the right and propagates along the two respective waveguides, and coupled into the two racetrack ring resonators as shown in Fig.1. The bus waveguides have the same cross-section dimensions as their respective racetrack ring resonators for good coupling.[26] The evanescent electric field of the two racetrack ring resonators provides an accelerating field in the central channel for sub-relativistic electrons entering from the left and exiting from the right. Fig. 1(b) shows the $E_x$ field distribution of a RRRA on resonance at one instant in time.

Considering electrons travelling in the mid-channel carrying negative charge, red indicates $E_x$ being accelerating field and blue indicates decelerating field. In addition, the two incident modes in the two waveguides are set to have a π phase difference to double the accelerating field $E_x$ in the channel as one can see from Fig. 1(b), and also cancels out the deflection force at the center of the channel caused by $E_y$ and $H_z$ as shown in Figs. 1(c) and 1(d). Since we carefully design the RRRA to have the field distributions in the channel traveling synchronously with the electrons at the speed of $v_e = c/n_{eff}$ ($n_{eff}$ is the effective refractive index of the waveguide), the co-propagating electrons will continue seeing this accelerating field distribution as they travel through the channel.

___

[a] Corresponding author: huiyang.deng@stanford.edu
[b] These two authors contribute equally to this work

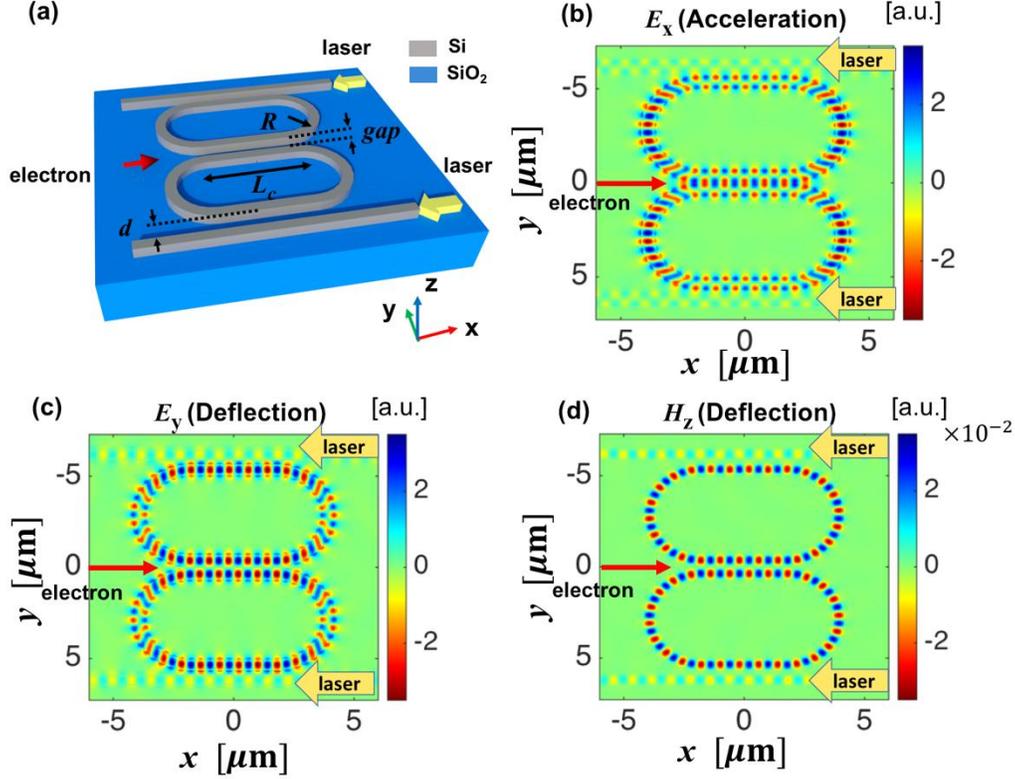

FIG. 1. (a) Schematic of a RRRA: Si waveguides and Si racetrack ring resonators on $SiO_2$ from a SOI substrate. (b) top view of $E_x$ electric-field profile (c) top view of $E_y$ electric-field profile (d) top view of $H_z$ magnetic-field profile.

Accelerator performance is usually characterized by the accelerating gradient $G$, the amount of energy gain transferred to the particle per unit distance travelled inside the accelerator, which is calculated as the average value of the real-time in-situ accelerating electric field $E_x[x(t), t]$ seen by the electrons during traveling through a total length of $L_{total}$ as shown in Equation (1).

$$G = <E_x> = \frac{1}{L_{total}} \int_0^{L_{total}} E_x[x(t), t] dx = F_A E_{inc} \qquad (1)$$

$L_{total}$ is the total span in $x$ direction of the RRRA ($L_{total} = L_c + 2R + Waveguide\ Width$, where $L_c$ is the straight part of the racetrack, and $R$ is the radius of the circular part of the racetrack). $F_A$, the field ratio, is a geometry dependent parameter to evaluate efficiency of energy transfer between the laser field and electrons.[7-10] $E_{inc}$ is the peak electric field amplitude of the incident laser.

The electrons will experience the highest accelerating field when amplitude of $E_x$ is maximized in the channel. However, the highest $E_x$ amplitude does not necessarily guarantee the highest accelerating gradient, because the gradient is the real-time in-situ integral of $E_x$ divided by $L_{total}$. This means that the ratio of effective accelerating length $L_{eff}$ (which is mainly the straight section $L_c$) over $L_{total}$ should not be too small, indicating $R$ cannot be too large. On the other hand, $R$ must not be too small. Otherwise, a sharp corner causes large bending loss, which leads to a smaller $E_x$. To maximize $G$ in Equation (1), we should find the optimal straight coupling length, $L_c$, ring radius, $R$, waveguide cross-section dimensions and the coupling distance $d$ between the waveguide and the straight coupling part of the racetrack.

The above-mentioned optimization criteria ensure potential to achieve maximum accelerating gradient $G$ in Equation (1). Nevertheless, due to the osscilation nature of laser field, electrons must be injected at the proper phase with respect to the oscillating field to actually experience the highest accelerating gradient. Fig. 2(a) shows one period of the accelerating force at one instant in time, where $x$-axis is a full $2\pi$ phase window (corresponding to a full effective wavelength in $x$ direction, as phase could be expressed as $\varphi = kx - \omega t_0$, where $k$ is the wave vector and $x$ is the distance in $x$ direction, with $\omega t_0$ being a constant in Fig. 2). The $y$-axis is the same as defined in Fig. 1. In Fig. 2(a), zero phase is the optimal phase condition, as it corresponds to the highest accelerating field. As the red color corresponds to positive accelerating force due to $E_x$, optical phase lying in between the two vertical dashed red lines is called the accelerating phase window. Being always inside the accelerating phase window, and locating close to the optimal phase condition will be optimal for acceleration.

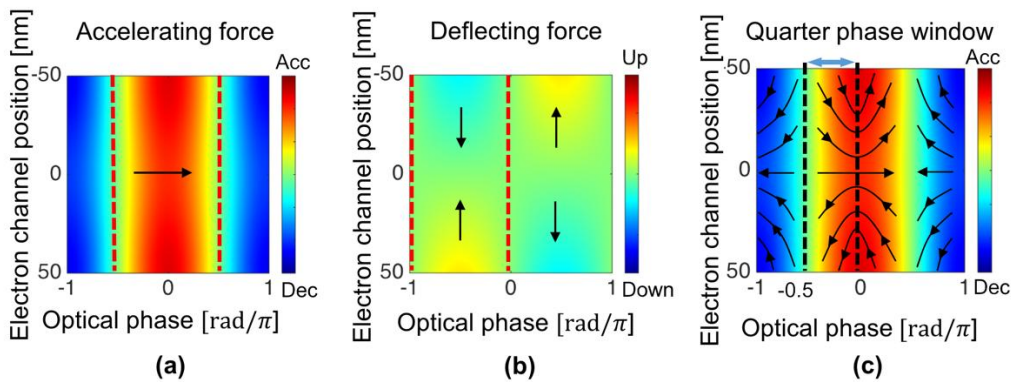

FIG. 2. (a) Accelerating force in the mid-channel caused by $E_x$. The optical phases in between the two vertical red dashed lines correspond to positive accelerating force, indicated by red in color bar, as well as the black arrow. (b) Deflecting force in the mid-channel caused by $-e(E_y + v_e B_z)$. The optical phases in between the two vertical red dashed lines correspond to self-focusing deflecting force, also indicated by black arrows. (c) The quarter phase window suitable for both sustained acceleration and controlled deflection is between the two black dashed vertical lines. The direction of total electromagnetic force is indicated by the black force vectors.

Furthermore, deflection control for an accelerator is always very important, and hence must be considered during RRRA optimization along with accelerating gradient optimization. As shown in Figs. 1(c) and 1(d), the deflection field is anti-symmetric. The forces associated with $H_z$ and $E_y$ are opposite in directions, thus will cancel out each other to some extent. Meanwhile, the deflection force of $E_y$ still dominates and determines the direction of the total deflection force, which is also proven mathematically in the supplementary material. The total deflection Lorentz force distribution is plotted in Fig. 2(b). Due to the anti-symmetry of $E_y$ and $H_z$, the deflection force is zero when the electrons propagate along the mid-line of the channel as in the ideal case. In reality, there always exist electrons that deviate from the mid-line. In that case, deflection can be controlled if the electrons are subject to a restoring force towards the mid-line. The deflecting force as a function of optical phase and channel position shown in Fig. 2(b) indicates that staying in the left half of the phase window means deflection in control.

To satisfy both criteria for sustained acceleration shown in Fig. 2(a) and controllable deflection shown in Fig. 2(b), a quarter phase window shown in Fig. 2(c) bounded between the two vertical black dashed lines is determined. In Fig. 2(c), the strength of accelerating force is indicated by the color bar, and the total electromagnetic force direction is indicated by the black force vectors. In this quarter phase window, the electrons will always experience sustained acceleration with controlled deflection.

We optimize the RRRA parameters under the condition that the optical phases are always in the quarter phase window within the entire duration of electrons travelling through the RRRA. (The detailed phase drift analysis is included in supplementary material). Here, we use an electron speed of 0.328c (corresponding to 30 keV, which is widely available from commercial electron guns) as a reasonable starting point. The optimization generates a RRRA design with $L_c$ = 4.08 μm, $R$ = 2 μm, $d$ = 50 nm, gap = 100 nm, waveguide width = 760 nm and height = 630 nm, which could accelerate 30 keV-electron to 34.135 keV over a length of $L_{total}$ = 8.84 μm. The peak incident electric field $E_{inc}$ is 448 MV/m, so that even the spot with the maximum electric field intensity within the structure still sees peak electric field amplitude below 2 GV/m, below the laser field induced damage threshold for silicon.[18] The average accelerating gradient of this one stage accelerator unit is 467.2 MeV/m, which is potentially more than twice the recent highest record for 30 keV-electron.[18] Meanwhile this design has a much lower demand for input electric field amplitude, as the field ratio $F_A$ is as high as 1.04, while previous designs for 30 keV electrons provide a very low $F_A$ (usually below 0.1).[8,15,18]

Based on the mechanism described above, we designed an accelerator system by cascading many RRRA units, starting with initial electrons speed of 0.328c (electron energy of 30 keV). The schematics is shown in Fig. 3(a). Within each stage, there is a RRRA unit with laser power coupled from the bus waveguides. In this accelerator system, the first stage is the RRRA we have described above. For the second stage, the initial speed of electron is 0.348c, i.e. the final speed of the first stage. We optimize parameters of the second RRRA based on previously discussed criteria. Again, there is a $F_A$ associated with this optimized RRRA geometry.[7-10] We determine the $E_{inc}$ by limiting every spot within the silicon structure sees electric field amplitude below 2 GV/m. With the optimized structures and $E_{inc}$, we finished design of second stage and achieved final speed in the second stage as 0.373c. We continue these optimization processes and finish the design of this 11-stage RRRA system with the parameters listed in Table I. The first two columns show the initial and final electron speed of each RRRA stage, denoted by $v_i$ and $v_f$. The effective wave speeds shown in Column 3 are increasing as the electrons are accelerated. Correspondingly, the waveguide cross-section dimensions listed in Column 4 and 5 are decreasing to match the requirement of decreasing $n_{eff}$. The sixth, seventh and eighth columns show the best geometry combinations of the RRRA to guarantee being on resonance and providing highest accelerating gradient. All field ratio $F_A$ values shown in Column 9 are above 1, indicating RRRA based DLA being very efficient utilizing the laser field. Low $E_{inc}$ values listed in Column 10 indicate lower input field demand for this design, due to the resonance effect. The tenth column shows the average gradient $G$, which is calculated as the product of $F_A$ and $E_{inc}$. $Q$ listed in Column 12 is monotonically decreasing due to the trend of decreasing waveguide dimensions, which means looser confine of electric field inside silicon. Cavity build-up time is hence also decreasing, as it scales linearly with $Q$. The acceleration performance of this RRRA system as a function of travel distance is plotted in Fig. 3(b), where the different colors correspond to different stages, and are matched to the color scheme of Fig. 3(a). This designed system accelerates electrons from 30 keV to 148.312 keV in 104.655 μm, with an overall average gradient of 1.130 GeV/m.

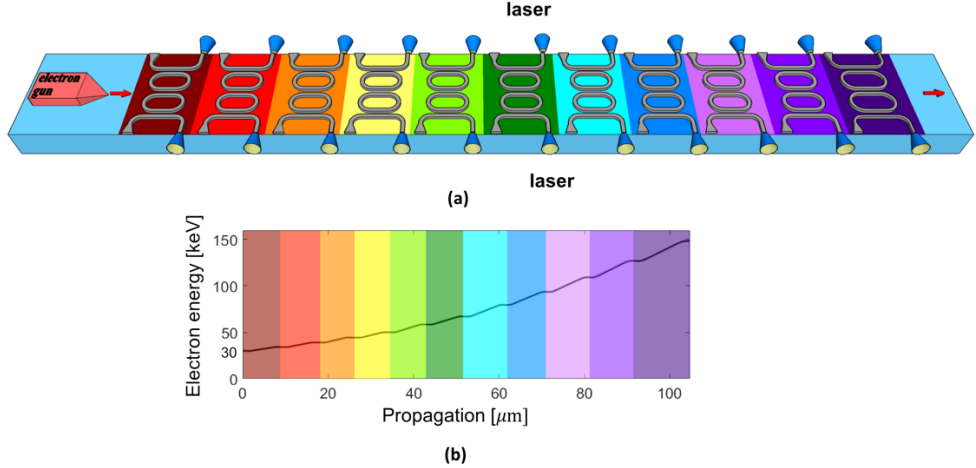

FIG. 3. (a) Schematic of the designed cascaded RRRA system (c) Acceleration energy gain as a function of propagating distance for the designed RRRA system

TABLE I. Parameters of designed cascaded RRRA systems

| $v_i$ | $v_f$ | Wave speed | Width (nm) | Height (nm) | $L_c$ (μm) | $R$ (μm) | $d$ (nm) | Field ratio | $E_{inc}$ (MeV/m) | $G$ (MeV/m) | $Q$ | Build up time (fs) |
|---|---|---|---|---|---|---|---|---|---|---|---|---|
| 0.328c | 0.348c | 0.337c | 760 | 630 | 4.08 | 2.00 | 50 | 1.04 | 448 | 467 | 5057 | 1809 |
| 0.348c | 0.373c | 0.358c | 645 | 540 | 4.04 | 2.00 | 90 | 1.54 | 372 | 572 | 4730 | 1798 |
| 0.373c | 0.397c | 0.381c | 575 | 475 | 3.47 | 2.15 | 140 | 1.81 | 350 | 633 | 4373 | 1767 |
| 0.397c | 0.413c | 0.402c | 535 | 435 | 3.45 | 2.20 | 180 | 1.27 | 533 | 679 | 3901 | 1665 |
| 0.413c | 0.442c | 0.427c | 500 | 400 | 4.27 | 2.00 | 280 | 2.39 | 402 | 962 | 3333 | 1511 |
| 0.442c | 0.468c | 0.456c | 475 | 365 | 4.49 | 2.00 | 340 | 2.44 | 393 | 960 | 2932 | 1420 |
| 0.468c | 0.502c | 0.483c | 455 | 345 | 5.07 | 2.00 | 400 | 2.31 | 571 | 1317 | 2791 | 1431 |
| 0.502c | 0.535c | 0.517c | 445 | 310 | 5.29 | 2.00 | 460 | 2.50 | 575 | 1439 | 2357 | 1293 |
| 0.535c | 0.567c | 0.550c | 440 | 280 | 5.42 | 2.00 | 520 | 2.20 | 724 | 1595 | 1900 | 1109 |
| 0.567c | 0.599c | 0.584c | 440 | 250 | 5.80 | 2.10 | 600 | 2.12 | 804 | 1707 | 1873 | 1160 |
| 0.599c | 0.632c | 0.616c | 425 | 240 | 7.68 | 2.50 | 680 | 1.60 | 1014 | 1626 | 1594 | 1043 |

In real experiment, it requires more thorough studies considering more limitations. In our current simulations, the end of the bus waveguide is terminated with perfect matching layer (PML) boundary conditions to prevent reflection at the abrupt waveguide air interface in the FDTD simulation. In reality, this could be done with impedance matching, such as utilizing tapered waveguide at the end, so that the gradual impedance change reduces reflection to prevent standing wave formation. In addition, due to the defects in fabrications of waveguides and ring resonators, the laser induced damage threshold could be lower than expected, which will decrease the accelerating gradient of our design. The fabrication defects could also cause unexpected losses and non-linear effects, etc. Besides, good coupling of the free-space laser onto the on-chip waveguide arrays is also an important topic that collaborating research groups are investigating. The lowest electron speed that this RRRA is useful is limited by the refractive index of waveguide core (Si in this case); the highest final electron speed that this RRRA could achieve is limited by the refractive index of substrate (SiO$_2$ in this case). So far, without further improvement ( such as using other materials or waveguide structures), this design is only useful for accelerating electrons from low-β (30 keV electron energy, β=0.328) to mid-β (148. 312 keV electron energy, β=0.632). Despite the limitations and challenges in experiments, this new idea still poses an alternative to realize an electron accelerator by manipulating electric field distribution with racetrack ring resonators.

In summary, we present a new design of racetrack ring resonator based monolithic laser-driven waveguide-coupled electron accelerator. This design provides a high accelerating field for electrons because of its resonance effect. The optimized 11-stage RRRA system is able to accelerate electrons from 30 keV to 148.312 keV in 104.655 μm with an average gradient of 1.130 GeV/m, which could potentially more than double the record of the highest gradient achieved so far for sub-relativistic acceleration.[14] In addition, this novel design has very high field ratio ($F_A > 1$, indicating efficient in utilizing the laser field, i.e. small laser input demand). Meanwhile, the design shows very good deflection control ability. The RRRA based DLA system could provide a realistic solution to fully integrate many DLA stages on-chip, showing potential to realize an "accelerator on a chip".

This work is supported by Gordon and Betty Moore Foundation (1186477-601-UAGZE) and Air Force Office of Scientific Research (AFOSR)(FA9550-14-1-0190). We thank the rest of the ACHIP team for helpful discussions.